\documentclass[preprint,showpacs,preprintnumbers,nofootinbib,amsmath,amssymb]{revtex4}

\usepackage{amssymb,color}
\usepackage{graphicx}
\usepackage{dcolumn}
\usepackage{bm}
\usepackage{subfigure}
\usepackage{hyperref}

\def\beqn{\begin{eqnarray}} 
\def\eeqn{\end{eqnarray}} 
\def\be{\begin{equation}}
\def\ee{\end{equation}}

\begin{document}

\title{Gravity Waves from a Cosmological Phase Transition:\\ Gauge Artifacts and Daisy Resummations}

\author{Carroll Wainwright}
\email{cwainwri@ucsc.edu} \affiliation{Department of Physics, University of California, 1156 High St., Santa Cruz, CA 95064, USA}
\author{Stefano Profumo}
\email{profumo@scipp.ucsc.edu} \affiliation{Department of Physics, University of California, 1156 High St., Santa Cruz, CA 95064, USA}\affiliation{Santa Cruz Institute for Particle Physics, Santa Cruz, CA 95064, USA} 
\author{Michael J. Ramsey-Musolf}
\email{mjrm@physics.wisc.edu} \affiliation{University of Wisconsin-Madison, Department of Physics
1150 University Avenue, Madison, WI 53706, USA}

\date{\today}

\begin{abstract}
\noindent The finite-temperature effective potential customarily employed to describe the physics of cosmological phase transitions often relies on specific gauge choices, and is manifestly not gauge-invariant at finite order in its perturbative expansion. As a result, quantities relevant for the calculation of the spectrum of stochastic gravity waves resulting from bubble collisions in first-order phase transitions are also not gauge-invariant. We assess the quantitative impact of this gauge-dependence  on key quantities entering predictions for gravity waves from first order cosmological phase transitions. We resort to a simple abelian Higgs model, and discuss the case of $R_\xi$ gauges. By comparing with results obtained using a gauge-invariant Hamiltonian formalism, we show that the choice of gauge can have a dramatic effect on theoretical predictions for the normalization and shape of the expected gravity wave spectrum. We also analyze the impact of resumming higher-order contributions as needed to maintain the validity of the perturbative expansion, and show that doing so can suppress the amplitude of the spectrum by an order of magnitude or more. We comment on open issues and possible strategies for carrying out \lq\lq daisy resummed" gauge invariant computations in non-Abelian models for which a gauge-invariant Hamiltonian formalism is not presently available. 
\end{abstract}


\preprint{NPAC-11-04}


\maketitle

\section{Introduction}
The search for gravitational waves is entering an exciting phase. The current generation of experiments is already delivering interesting  results, including recent limits on the amplitude of stochastic gravitational wave backgrounds from the LIGO Collaboration \cite{ligobckg,pta}. Rapid advances in the development of space-borne detectors \cite{lisapath} that might be operational in the relatively near future are also expected \cite{decadal}.

As pointed out long ago \cite{earlypt}, cosmological phase transitions in the early universe might produce an imprint  in the form of a stochastic background of relic gravity waves. These would arise as a result of the collision or turbulent motion of bubbles of ``true vacuum'' expanding and eventually filling the metastable vacuum in a cosmological first order phase transition \cite{kkt}. The resulting signal might be large enough to be detectable by the next generation gravity wave search experiments, providing a unique window on the  early cosmological history of the universe \cite{gwdetection}.

The spectrum of the gravity wave background arising from a first order cosmological phase transition is controlled by two physical properties of the phase transition itself: (a) the amount of false vacuum energy liberated during the phase transition -- in other words, the latent heat associated with the transition; and (b) the bubble nucleation time scale, which gives a measure of how rapidly the phase transition occurs relative to the early universe Hubble expansion rate \cite{kkt,shapo}. The resulting gravity wave normalization and spectral peak can be estimated as a function of these two physical quantities. Detailed analytical \cite{kkt, apreda} and numerical \cite{konstadin, caprini} studies exist that relate the two parameters to the predicted spectrum, in particular for the case of detonations, where the speed of the bubble wall is larger than the speed of sound (see Ref.~\cite{caprini} for a discussion of the opposite case of deflagration).

One class of models where a strongly first order phase transition is a necessary ingredient is electroweak baryogenesis \cite{kuzminrubakovshapo}. In the presence of B-violating electroweak sphalerons in the Standard Model(SM)  and sources of CP-violation beyond those of the CKM CP violating phase, the electroweak phase transition (EWPT) can produce a sufficiently large baryon  number density to explain the observed baryon asymmetry (for recent studies see e.g. \cite{Carena:1996wj, Huber:2001xf,  Kang:2004pp, Carena:2008vj,  Funakubo:2009eg, Chiang:2009fs, Chung:2010cd, Profumo:2007wc, Ham:2010ha,  Lee:2004we, Chung:2008ay, Cirigliano:2009yd, Chung:2009qs}. To prevent the washout of the produced asymmetry, the phase transition must be strongly first order, thus necessarily producing gravity waves. Interestingly, the typical frequencies at which gravity waves from the EWPT are red-shifted today falls in the milli-Hertz to Hertz range, where the expected sensitivity of the space-based interferometer LISA is maximal. The search for a gravity wave relic from the EWPT is therefore especially intriguing and promising (see e.g.\ \cite{Maggiore:2000gv, Kehayias:2009tn}).

From a particle physics perspective, the determination of the details of an EWPT
depends on the calculation of the  finite-temperature effective action $\Gamma_\mathrm{eff}$ as a function of the background bosonic fields --- denoted generically here as $\varphi(x)$ --- that are present in the theory. In the case of the SM, lattice computations and the LEP lower bound on the mass of the Higgs boson  imply that electroweak symmetry breaking in a SM universe occurs through a cross-over transition\cite{Kajantie:1996mn,Aoki:1999fi}. To obtain a strongly first-order EWPT as needed for both electroweak baryogenesis (EWB) and the associated relic gravity waves, one must augment the SM scalar sector by the addition of new scalar fields, such as a second Higgs SU(2$)_L$ doublet as in the Minimal Supersymmetric Standard Model (MSSM) \cite{Carena:1996wj, Huber:2001xf,  Kang:2004pp, Carena:2008vj,  Funakubo:2009eg, Chiang:2009fs, Chung:2010cd} or a real singlet in minimal extensions of the SM scalar sector (see e.g. \cite{Profumo:2007wc}). Electroweak bubble nucleation occurs when a combination of the one of these new fields and the neutral component of the SM Higgs doublet  becomes non-zero. The properties of relic gravity waves produced by bubble collisions then follows from the behavior of $\Gamma_\mathrm{eff}[\varphi(x)]$. 

The most theoretically robust computations of  the effective action are performed using non-pertrubative (lattice) methods. Given the cost of such computations, however, this approach is not feasible for exploring EWPT dynamics in a variety of beyond the Standard Model (BSM) scenarios. Consequently, one must resort to perturbation theory which, in turn, requires introduction of gauge-fixing. 
As recognized long ago \cite{jackdolan}, perturbative computations of the effective potential --- and more generally $\Gamma_\mathrm{eff}[\varphi(x)]$ ---  generically lead to a gauge-dependent function. Physical quantities like the latent heat or the bubble nucleation rate should not, however, exhibit any gauge dependence.

In fact, general arguments  imply that the critical temperature \cite{jackdolan2} and the bubble nucleation rate \cite{bubbles} are gauge independent. These statements follow from the so-called Nielsen identities and their generalization \cite{nielsen,followupnielsen} that describe the dependence of the effective action on the gauge fixing condition imposed on the quantized fields. In particular, the effective action $\Gamma_\mathrm{eff}[\varphi(x)]$ is gauge-invariant when $\varphi(x)$ is an extremal configuration, that is, one satisfying the equations of motion\footnote{In the case of the effective potential, $\varphi(x)=\mathrm{const}\equiv\varphi_\mathrm{min}$ is just the spacetime independent background field ({\em e.g.}, Higgs vacuum expectation value) that gives a minimum or maximum of the potential.}.  In principle, then, one should be able to obtain physical, gauge-invariant quantities relevant to the EWPT from $\Gamma_\mathrm{eff}$ by working with an appropriate set of extremal field configurations.  

In practice, a non-trivial gauge-dependence can enter perturbative computations from an inconsistent truncation of the perturbative expansion\cite{mjrmpatel}. In the context of sphaleron-induced baryon number washout, following on earlier work by Refs.~\cite{Laine:1994zq,Baacke:1993aj,Baacke:1994ix}, it was shown in Ref.~\cite{mjrmpatel} how a consistent, systematic order-by-order approach can yield a gauge-invariant perturbative result. However, most of the remaining literature on the topic of the EWPT, including the context of gravity wave production and of baryogenesis relevant here, appears to suffer from gauge-dependence (typically, perturbative calculations of EWPT-relevant quantities have been performed in the Landau gauge)\footnote{To our knowledge, there exist no non-perturbative computations of all of the quantities relevant for predictions of GW spectra.}. Apart from the point of principle, the question then arises as to the quantitative impact that this gauge artifact has upon predictions of observable quantities.

In the present study, we address this question as it pertains to computations of gravitational wave spectra from a first order EWPT. To that end, we consider the simplest  model involving scalar fields charged under a gauge group: the Abelian Higgs model, also known as the Coleman-Weinberg or scalar QED model. We then resort to a class of gauges known as $R_\xi$ (or  renormalizable) gauges, and we calculate the effective potential at finite temperature including its explicit dependence on the parameter $\xi$. 

Studying the Abelian Higgs model has two clear advantages. First, its parameter space is small and easily analyzed. Second, and more importantly, one may calculate its effective potential using a  gauge-invariant  Hamiltonian approach\cite{boyan} whose results can be compared with those obtained from  the gauge-dependent approach. 
We then calculate quantities relevant to the character, strength and duration of the EWPT both in the $R_\xi$ gauges (including Landau gauge) and in the gauge-invariant approach, and systematically compare the results. 

We find that the gauge choice may have a dramatic impact, amounting to several orders of magnitude, on the inferred gravity wave spectrum, and even on the first- or second-order character of the phase transition itself. We also observe that the Landau gauge results  closely matches, at least for the Abelian Higgs model, the results using the explicitly  gauge-invariant Hamiltonian formulation. This situation is perhaps not surprising, given the arguments in Ref.~\cite{fischler} (see below). We caution however that this conclusion might not be easily generalizable to non-Abelian gauge theories and that even in the Abelian Higgs model, a gauge-invariant resummation of higher-order terms that would otherwise spoil the convergence of the perturbative expansion remains to be developed\footnote{A gauge-invariant prescription for estimating these terms was developed in Ref.~\cite{mjrmpatel}.}. To underscore the importance of the latter problem, we study the impact of including the \lq\lq daisy resummation"  in $R_\xi$ gauges. We find that in the Landau gauge, inclusion of the resummation typically reduces the overall amplitude of the GW spectrum compared to the gauge-invariant but un-resummed result. 
We then comment on strategies to tackle these issues in non-Abelian models (such as the Standard Model or its supersymmetric extensions) -- including those of Ref.~\cite{mjrmpatel}.

We begin in section~\ref{sec:gaugeDep} with the explicit calculation of the gauge-dependence of the effective potential, at both zero and finite temperature, for the Abelian Higgs model. In section~\ref{sec:GWparams}, we explain the calculations required to predict gravitational wave spectra and other physical observables related to the phase transition. Finally, we present our results and conclusions in sections~\ref{sec:results} and~\ref{sec:discussion}.

\section{Gauge Dependence of the Effective Potential}
\label{sec:gaugeDep}
We are concerned here with an Abelian Higgs model encompassing a complex singlet Higgs field with Lagrangian
\begin{equation}
\mathcal{L} = -\frac{1}{4}F_{\mu\nu}F^{\mu\nu} + \frac{1}{2}(D_\mu \phi)^*D^\mu \phi - V_0(\phi^*\phi),
\end{equation}
where $F_{\mu\nu} = \partial_\mu A_\nu - \partial_\nu A_\mu$ and $D_\mu = \partial_\mu - ieA_\mu$ are the standard electromagnetic tensor and covariant derivative, respectively. The potential $V_0$ is
\begin{equation}
V_0(\phi^*\phi) = -\frac{1}{2}m^2\phi^*\phi + \frac{1}{8}\lambda (\phi^*\phi)^2.
\end{equation}
The tree-level vacuum expectation value is $v^2  = \frac{2m^2}{\lambda}$, and the bare Higgs mass at the vacuum expectation value (vev) is $m_h^2 = 2m^2$.

In order to perform perturbative calculations, we must add gauge-fixing and ghost terms to the Lagrangian. In the $R_\xi$ gauge, these are
\begin{equation}
\label{eq:gfghost}
\mathcal{L}_{gf} + \mathcal{L}_{ghost} = -\frac{1}{2\xi}(\partial_\mu A^\mu + \xi ev\eta)^2 + \partial_\mu \bar{c} \partial^\mu c  - \xi e^2 v \sigma \bar{c}{c},
\end{equation}
where $c$ is the Grassmann-valued ghost field, and where we have split $\phi$ into its real and imaginary components: $\phi = \sigma + i\eta$. We choose the vev such that $\langle\eta\rangle = 0$, making $\sigma$ the physical Higgs boson and $\eta$ the non-physical goldstone boson. Note that in Landau gauge ($\xi = 0 $, fixing $ \partial_\mu A^\mu = 0$), the ghost field completely decouples from the theory.

To include finite-temperature corrections, we must go to (at least) one-loop order in the effective potential. At zero-temperature, the calculation of the one-loop effective potential proceeds by taking the trace of the inverse propagators for each particle. This yields terms like $\frac{1}{2}\int\frac{d^4k}{(2\pi)^4} \log (k^2 +m_i^2(\sigma))$, although determining the proper expression for the $\xi$-dependence of the gauge boson is somewhat complicated by the need to sum over Lorentz indices ( see, {\em e.g.}, Ref.~\cite{Delaunay:2007wb}). Using {$\overline{MS}$} renormalization, the full one-loop zero-temperature potential is
\begin{equation}
\label{eq:V1}
V_1(\sigma, T\!=\!0) = \sum_{\text{particles}}\frac{n_i}{64\pi^2} m_i^4(\sigma)\left[\log\left(\frac{m_i^2(\sigma)}{\mu^2} \right) - c\right],
\end{equation}
where $n_i$ are the degrees of freedom for each particle and $c = 1$ for the gauge boson's transverse modes and $c=\frac{3}{2}$ for its other modes and all other particles\footnote{In the literature, authors generally use $c=\frac{5}{6}$ for all three physical modes of the gauge boson. This only makes a difference if one includes thermal masses in the zero-temperature potential, which is a small correction that few authors include.}. Table~\ref{tab:particles} lists all particle masses and their degrees of freedom. Several of the masses are gauge-dependent, and, precisely because of this fact, the effective potential is also gauge-dependent. Note that at both the origin and the tree-level vev ($\sigma^2 = v^2 = 2m^2/\lambda$) the gauge-dependence disappears\cite{mjrmpatel}, as expected from the Nielsen identities \cite{nielsen,followupnielsen}. However, the value of $\sigma$ that minimizes the one-loop effective potential {\em is not} gauge invariant.

For the particular case of the Abelian Higgs model, Fischler and Brout \cite{fischler} defined an effective potential from the vacuum-to-vacuum $S$-matrix element  without resorting to the introduction of sources, a procedure that contrasts with the conventional definition in terms of the Legendre transform of the source-dependent generating functional, $Z[j]$. In this context, the \lq\lq free-energy" is minimized by a spacetime-independent background field only in the Landau gauge,  whereas in other gauges the minimizing fields must carry a spacetime dependence. Consequently, only in the Landau gauge does the minimum of the effective potential in the absence of sources characterize the presence or absence of symmetry-breaking. For the formulation with sources, a spacetime-independent background field will yield the minimum of energy in any gauge of the form given  in  Eq.~(\ref{eq:gfghost}), implying equal values of the minima of the effective potential for any choice of $\xi$  \cite{nielsen,followupnielsen}. While the formulation of Ref.~\cite{fischler} is manifestly gauge invariant by construction, its relationship with the development in terms of sources has not to our knowledge been clarified. That  being said, the arguments of Ref.~\cite{fischler} are suggestive that results obtained with the Landau gauge effective potential may be most physically reasonable.   Indeed, we find close numerical agreement between Landau gauge quantities and those obtained using an explicitly gauge-invariant Hamiltonian formalism (see below). We emphasize, however, that this agreement does not carry over to the non-Abelian case.

\begin{table}
\begin{tabular}{r | c | c | c}
 particles & d.o.f. & (masses)$^2$  & (thermal masses)$^2$ \\ \hline \hline
 transverse gauge polarization & 2 & $e^2\sigma^2$ & \\ \hline
 longitudinal gauge polarization & 1 & $e^2\sigma^2$ & $\frac{1}{3}e^2T^2$ \\ \hline
 time-like gauge polarization & 1 & $\xi e^2\sigma^2$ & \\ \hline
 higgs boson & 1 & $-m^2 + \frac{3}{2}\lambda \sigma^2$ & $(\frac{1}{3}\lambda + \frac{1}{4}e^2)T^2$ \\ \hline
 goldstone boson & 1 & $-m^2 + (\frac{1}{2}\lambda +\xi e^2)\sigma^2$ & $(\frac{1}{3}\lambda + \frac{1}{4}e^2)T^2$ \\ \hline
 ghosts & -2 & $\xi e^2\sigma^2$ & \\ 
\end{tabular}
 \caption{\label{tab:particles} Particle content of the Abelian Higgs model, including Fadeev-Popov ghosts. One ghost effectively cancels the contribution from the unphysical time-like polarization, while the other cancels either the longitudinal polarization (at $\sigma=0$) or the goldstone boson (at $\sigma=v$).}
 \end{table}

The finite temperature contribution can be derived similarly to the zero-temperature contribution, except that the integral over momenta is replaced with a sum over Matsubara modes: $\int dk^0 \rightarrow \frac{1}{\beta}\sum_\beta$ and $k^0 \rightarrow \frac{2n\pi}{\beta}$. This yields
\begin{equation}
V_1(\sigma, T > 0) = \frac{T^4}{2\pi^2} \sum_\text{particles} J\left[\frac{m_i^2(\sigma)}{T^2}\right],
\end{equation}
where
\begin{equation}
J(x^2) \equiv \int_0^\infty dy\; y^2 \log \left(1- e^{-\sqrt{y^2+x^2}}\right).
\end{equation}
In the high-temperature (low-$x$) limit,
\begin{equation}
\label{eq:high}
J(x^2) \approx -\frac{\pi^4}{45} + \frac{\pi^2}{12}x^2 - \frac{\pi}{6}x^3 - \frac{x^4}{32}\log\frac{x^2}{a_b} - \mathcal{O}(x^6)
\end{equation}
where $\log a_b = \tfrac{3}{2} - 2\gamma_E + 2\log(4\pi)$ and $\gamma_E$ is the Euler constant \cite{Anderson:1992}. All higher-order terms are simple polynomials in $x^2$. Again, the gauge dependence disappears at $\sigma^2 = v^2$ and at $\sigma = 0$, but it is non-trivial everywhere else. We also observe that the general arguments in Ref.~\cite{fischler} do not depend on whether one works with a Minkowski or Euclidean  formulation of the functional integral appearing in the generating functional, so that even at finite-$T$ use of the Landau gauge is equivalent to a gauge-invariant formulation for the Abelian Higgs model.

\subsection{Thermal Mass Corrections}
It is well-known that near the critical temperature for a phase transition, validity of the perturbative expansion of the effective potential breaks down. Quadratically divergent contributions from non-zero Matsubara modes must be re-summed through inclusion of thermal masses in the one-loop propagators\cite{gross1981,parwani1992}: $m^2(\sigma) \rightarrow m^2_{eff}(\sigma) = m^2(\sigma)+m^2_{therm}(T)$. Table~\ref{tab:particles} lists all thermal mass corrections (see ref.~\cite{arnold1993} for further discussion and explicit calculations of the masses). Generally, one performs this \lq\lq daisy resummation" by only including the thermal masses in the zero-mode propagators, which results in a mass correction to only the cubic term in the effective potential. It is slightly more convenient from a computational standpoint to include the corrections in all propagators, although we do check that this only makes a small difference in the resulting potential.

\subsection{Alternative Gauge Invariant Formulation}
For comparison, we also examine the gauge-invariant effective potential put forward by Boyanovsky et al.\ \cite{boyan}. These authors derive the potential by working in the Hamiltonian formalism using only gauge-invariant physical states. In this case, there exist only four independent degrees of freedom (two transverse gauge, one longitudinal gauge, and Higgs), with no need for ghost cancellations. The unrenormalized one-loop effective potential is
\begin{equation}
V_1(\chi) = \frac{1}{2}\int\frac{d^3k}{(2\pi^3)}\left[ 2\omega_T + \omega_h + \omega_p \right],
\end{equation}
where $\omega_T^2 = k^2 + m_T^2$ and $\omega_h^2 = k^2+m_h^2$ arise from the transverse gauge and Higgs degrees of freedom, respectively, and the plasma frequency $\omega_p^2 = (k^2+m_g^2)(k^2+m_T^2)/k^2$ contains the contribution of both the gauge boson's longitudinal polarization and the Goldstone boson. The order parameter $\chi$ is a spacetime-independent, gauge-independent shift of the field, and the gauge, Higgs, and Goldstone masses $m_T$, $m_h$, and $m_g$ are given in table~\ref{tab:particles} with $\xi=0$ and $\sigma \rightarrow \chi$. The tree-level potential is unchanged.

The first two contributions to $V_1(\chi)$ exactly match the transverse gauge polarization and Higgs boson contributions to the potential in $R_\xi$ gauge. At the tree-level extrema, the plasma frequency also matches the contributions from all other modes.  However, away from the tree-level extrema, the plasma frequency does not match and looks similar only to Landau gauge ($\xi=0$), as one would expect from the general arguments in Ref.~\cite{fischler}.  
Therefore, we anticipate the Landau gauge will provide a close approximation to the gauge-independent Hamiltonian result, a conclusion similar to what was found in Boyanovsky et al. \cite{boyan}. 

Using $\overline{MS}$ regularization (see the appendix), we find that the plasma frequency contribution to the one-loop potential is
\begin{align}
V_{1p}(\chi, T\!=\!0) =&  \frac{1}{64\pi^2} \left[ (m_T^2-m_g^2)^2  \left(\log\frac{m_T^2-m_g^2}{\mu^2} - \frac{3}{2}  \right) + 4m_T^2 m_g^2 \right] \\
V_{1p}(\chi, T\!>\!0) =&\; \frac{T^4}{2\pi^2} J_2\left(\frac{m_T^2}{T^2}, \frac{m_g^2}{T^2} \right),
\end{align}
where $J_2$ is calculated by Boyanovsky et al.\ to be
\begin{equation}
J_2(a^2, b^2) \equiv \int_0^\infty dy\; y^2 \log \left[1- e^{-\frac{1}{x}\sqrt{(y^2+a^2)(y^2+b^2)}}\right].
\end{equation}
In the high temperature expansion, Boyanovsky et al.\ find that their gauge invariant potential is the same as the potential in Landau gauge up to the cubic terms, but the equality breaks down beyond this.

In what follows, we compare results for the GW spectra using the Lagrangian and gauge-invariant Hamiltonian methods. We observe that the daisy resummation of higher-order contributions was not considered in Ref.~\cite{boyan}, and it is not immediately clear how one would do so. Consequently, when comparing results in the two approaches, we will not include the Daisy resummation in the Lagrangian formulation.

\section{Calculation of GW Parameters}
\label{sec:GWparams}
We calculate several parameters of interest to gravitational wave production from early universe phase transitions using the Abelian Higgs model with full and explicit gauge dependence in the class of $R_\xi$ gauges. These include the transition temperature $T_*$, the minima of the low and high-temperature phases at the transition, the relative change in energy density $\alpha$, and the approximate duration of the phase transition $\beta^{-1}$.

\subsection{Calculating the Transition Temperature}
In first-order cosmological phase transitions, the low-temperature phase develops by nucleating bubbles within the high-temperature phase (see Refs.~\cite{coleman,linde}  for original work on cosmological transitions). A critical bubble---one whose surface tension exactly balances its outward pressure---is given by the $\mathcal{O}(3)$ symmetric action
\begin{equation}
\label{eq:S3}
S_3 = 4\pi \int_0^\infty r^2dr\left[ \frac{1}{2}\left(\frac{d\sigma}{dr}\right)^2+V(\sigma(r),T)\right],
\end{equation}
subject to the constraints that the field is smooth at $r=0$ and in the high-temperature minimum at $r=\infty$. Smaller bubbles collapse, while larger bubbles grow and eventually fill the universe with the new phase. Equation~\ref{eq:S3} yields the radial equation of motion
\begin{equation}
\label{eq:euclid}
\frac{d^2\sigma}{dr^2} + \frac{2}{r}\frac{d\sigma}{dr} = \frac{\partial}{\partial\sigma}V(\sigma,T),
\end{equation}
which we solve using the `undershoot/overshoot' method (see e.g.\ Ref.~\cite{apreda}). 

To find the exact transition temperature $T_\ast$, we must determine when the low-temperature phase nucleates at least one bubble per Hubble volume. The nucleation rate goes roughly as $\Gamma \propto T^4e^{-S_3/T}$, where the constant of proportionality can be found largely on dimensional grounds. For electroweak scales, this gives a transition temperature determined by $S_3/T_* \sim 140$ (see e.g.\ Ref.~\cite{Quiros:1999jp}). Note that the exponent changes very rapidly, so determining the exact form of the coefficient is quantitatively unimportant.

Finding the minima of the low- and high-temperature phases can be a non-trivial task, especially since the high-temperature minimum is not necessarily at $\sigma = 0$ and intermediate minima can develop for $\xi > 0$ (see section~\ref{sec:results} below). Our strategy, however, is fairly straightforward. We first observe that the transition occurs in the range $T_C > T_\ast > T_\mathrm{min}$, where $T_\mathrm{min}$ is the lowest temperature at which the original, high-temperature phase exists and $T_C$ is the temperature at which the minima of the potential in the two phases are degenerate.
We then trace the low-temperature minimum upwards from $T=0$ and the high-temperature minimum downwards from $T=T_0$ by numerically integrating 
\begin{equation}
\frac{d\sigma_{min}}{dT} = -\left( \frac{\partial^2 V}{\partial \sigma\partial T}\right) / \left(\frac{\partial^2 V}{\partial \sigma^2} \right).
\end{equation}
At each point, we calculate the transition rate by finding $S_3$. Following the evolution of the minima and $S_c$ we then  determine the temperature at which $S_3/T = 140$.

\subsection{Calculating the Latent Heat and Transition Duration}
With the transition temperature in hand, the relative change in energy densities $\alpha$ and transition duration $\beta^{-1}$ follow without much effort. 
When evaluated at its minimum, the effective potential is the same as the free energy density of the system\footnote{Here, we neglect kinetic energy contributions associated with non-vanishing gradients of the background field in the bubble walls.}. Therefore, the energy density difference between the two phases is
\begin{equation}
\Delta \rho  =[V(\sigma_{hot}, T_*)+s_{hot}T_*] - [V(\sigma_{cold}, T_*)+s_{cold}T_*],
\end{equation}
where the entropy density is $s = -\partial V/\partial T$. Note that at $T_*=T_c$, $\Delta \rho$ is identical to the latent heat. The quantity of interest in the production of gravitational waves is $\alpha = \Delta \rho/\rho_{rad}$, where $\rho_{rad} = \frac{g_*\pi^2}{30}T^4$ and $g_*$ is the number of relativistic degrees of freedom at the phase transition, which we take to be 100.

Writing the bubble nucleation rate as $\Gamma = \Gamma_0 e^{\beta t}$, $\beta^{-1}$ gives the approximate phase transition duration.
For a radiation dominated universe,
\begin{equation}
\frac{\beta}{H_*} = T_* \left.\frac{d(S_3/T)}{dT}\right|_{T_*}
\end{equation}
where $H_*$ is the Hubble expansion rate during the transition (see e.g. Ref.~\cite{apreda}).

\subsection{Calculating GW Spectra}
We employ here the analytical approximation provided in ref.~\cite{konstadin} to the numerical simulations carried out in that same work. We refer the Reader to ref. \cite{caprini} for further insights on the results of ref.~\cite{konstadin} The gravity wave spectrum (more precisely, the gravity wave energy density per frequency octave) from collisions at production is parameterized by
\begin{equation}
\Omega_{\rm GW* }(f_*)=\tilde\Omega_{\rm GW*}\frac{(a+b)\tilde f_*^bf_*^a}{b\tilde f_*^{(a+b)}+af_*^{(a+b)}},
\end{equation}
where the two exponents, obtained from fits to the numerical results, are set to $a=2.8$ and $b=1.0$. The spectrum is redshifted according to
\begin{equation}
\tilde f=16.5\times 10^{-3}\ {\rm mHz}\left(\frac{\tilde f_*}{\beta}\right)\left(\frac{\beta}{H_*}\right)\left(\frac{T_*}{100\ {\rm GeV}}\right)\left(\frac{g_*}{100}\right)^{1/6},
\end{equation} 
\begin{eqnarray}
h^2\tilde\Omega_{\rm GW} & = &1.67\times 10^{-5}\  \tilde\Omega_{\rm GW*} \left(\frac{100}{g_*}\right)^{1/3}\\
&=& 1.67\times 10^{-5}\tilde\Delta \ \kappa^2 \left(\frac{H}{\beta}\right)^2\left(\frac{\alpha}{\alpha+1}\right)^2\left(\frac{100}{g_*}\right)^{1/3},
\end{eqnarray} 
with the functions $(\tilde f_*/\beta)$ and $\tilde \Delta$ depending on the bubble wall velocity $v_b$ (and hence implicitly on the relative energy density difference $\alpha$) according to the following parameterization (again as given in ref.~\cite{konstadin}):
\begin{equation}
\tilde\Delta(v_b)=\frac{0.11\ v_b^3}{0.42+v_b^2}
\end{equation}
\begin{equation}
(\tilde f_*/\beta) (v_b)=\frac{0.62}{1.8-0.1v_b+v_b^2}.
\end{equation}
Finally, we employ the following parameterization for the bubble wall velocity \cite{Laine:1993ey}
\begin{equation}
v_b=\frac{\sqrt{1/3}+\sqrt{\alpha^2+2\alpha/3}}{1+\alpha},
\end{equation}
valid in the limit of interest here of strongly first order phase transitions. Note that the overall amplitude scales as $h^2\tilde\Omega_{\rm GW} \propto g_*^{-7/3}$ for $\alpha \ll 1$, so it can be changed by several orders of magnitude by choosing a model with a different $g_*$.

\section{Numerical Results}
\label{sec:results}

In the most basic model without any additional fields, four parameters determine the effective potential: the tree-level Higgs mass $m_h$, the tree-level vev $v$, the gauge coupling $e^2$, and the renormalization scale $\mu$. We vary only the first two of these, keeping $v = 246$ GeV and $\mu = 1$ TeV fixed. We include phase transition calculations using the gauge-invariant Hamiltonian formalism of Boyanovsky et al. \cite{boyan} without resummation, shown in the figures as solid arrows.

In order to generate a fairly strong first-order phase transition, the gauge boson mass must be relatively large. However, if the mass is too large then the one-loop zero-temperature potential overwhelms the tree-level potential and perturbation theory is unreliable. At the tree-level vev, $V_0(\sigma\!=\!v)=-\frac{1}{8}\lambda v^4$ and $V_1(\sigma\!=\!v, T\!=\!0) = \frac{3}{64\pi^2}(e^2v^2)^2[\log(\frac{e^2v^2}{\mu^2})-\frac{5}{6}]$ plus a small contribution from the Higgs boson. These two are roughly equal when $e^4 \approx 4\lambda$ or $e^2 \approx \frac{2m_h}{v}$. To be slightly more conservative, we demand that $e^2 \leq \frac{m_h}{v}$.

\begin{figure}[h]
	\centering
	\subfigure{
		\includegraphics[scale=1]{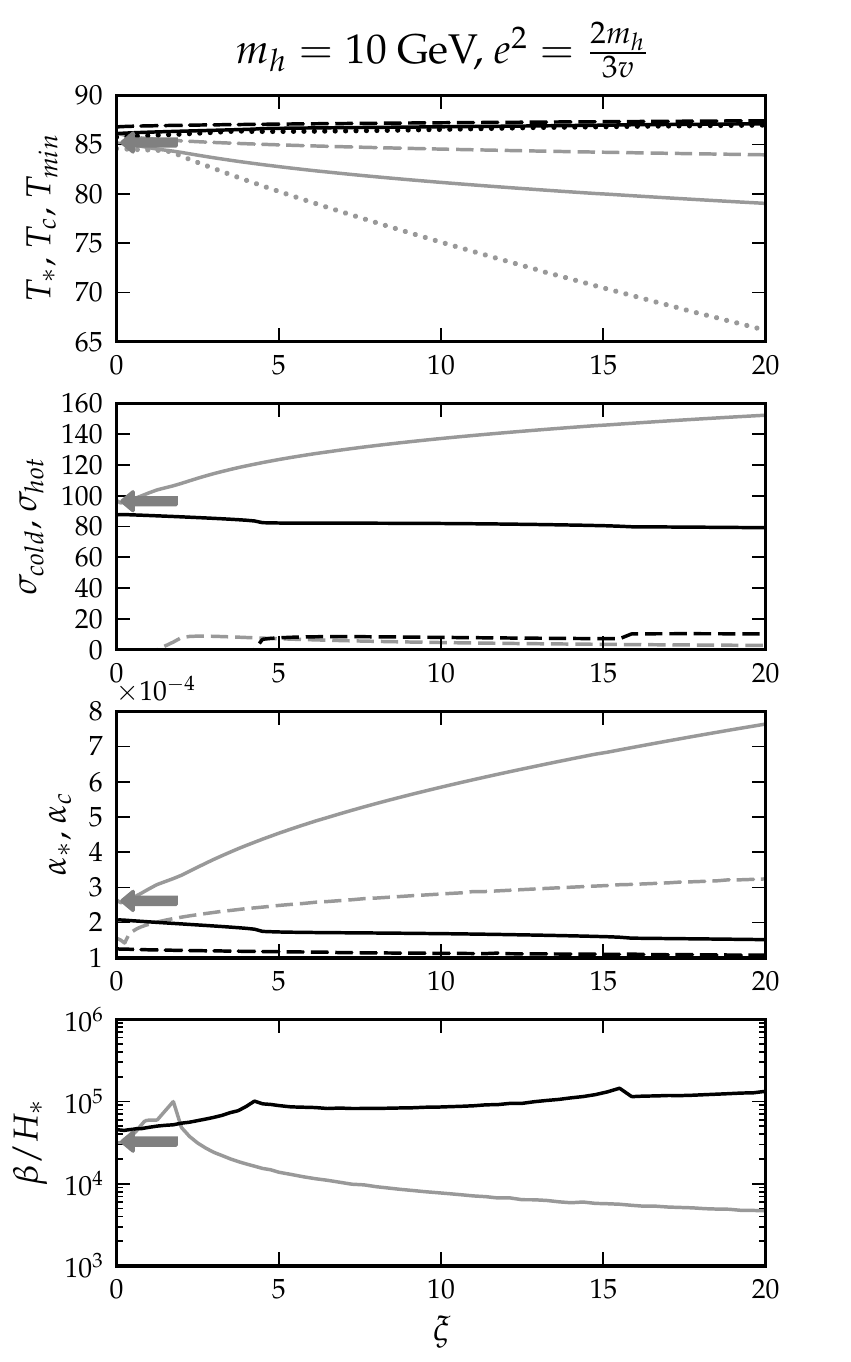}
		\includegraphics[scale=1]{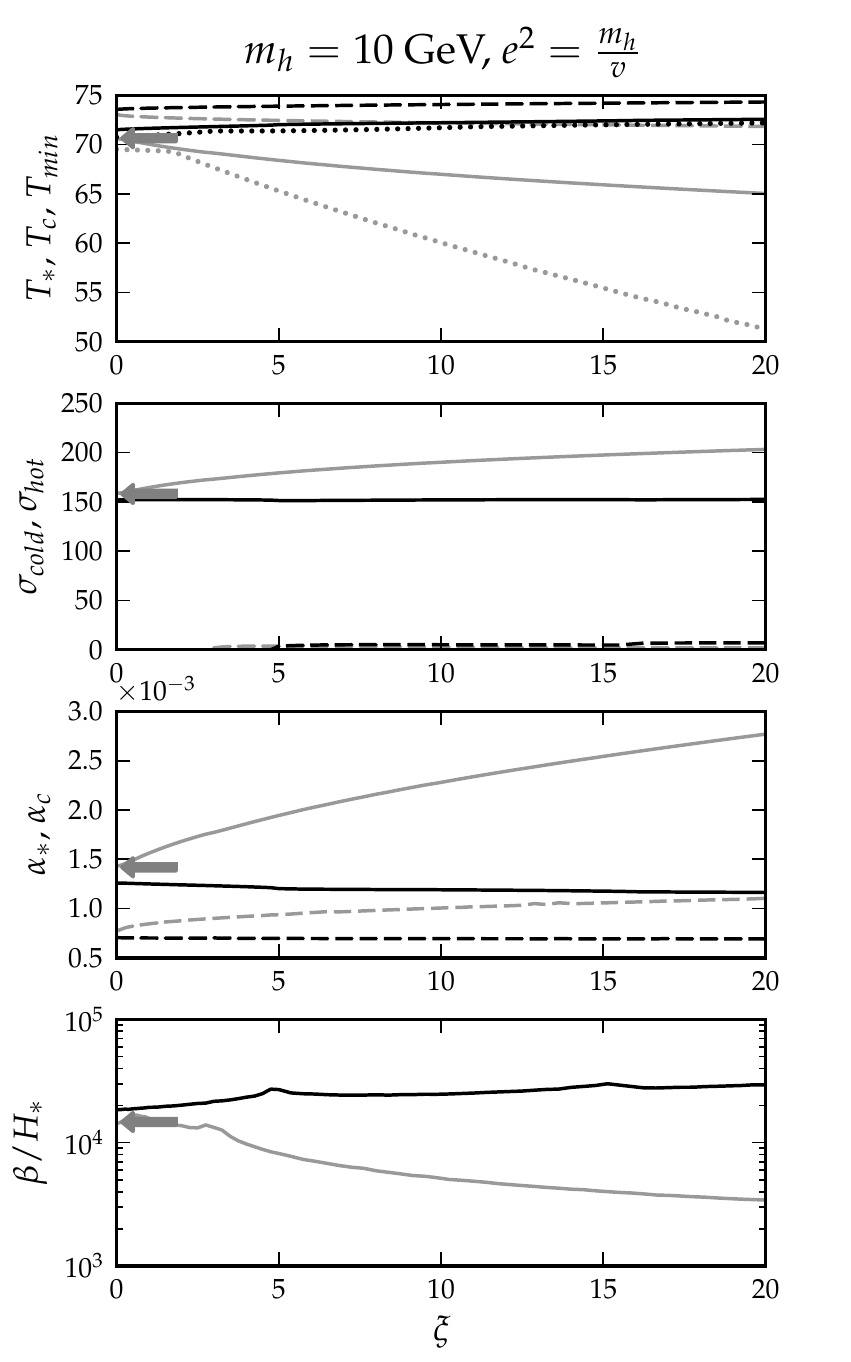}
	}
	\caption{\label{fig:10mh} Calculated gauge dependence of phase transition parameters for a low-mass Higgs boson. In all panels, black (grey) lines denote models with (without) resummation. The arrows denote values corresponding to the solid lines, but calculated in the gauge-invariant Hamiltonian formalism. All quantities along the y-axes are in units of GeV, except for $\beta/H$ which is unitless. In the first panel, solid, dashed and dotted lines denote the transition temperature $T_*$, the critical temperature $T_c$, and the minimum temperature at which the hot phase exists. In the second panel, solid and dashed lines denote the minima of the cold and hot phases. The third panel shows the relative difference in energy densities at both the critical temperature (dashed line) and the actual transition temperature (solid line). The final panel gives $\beta/H_*$, where $\beta$ is the approximate inverse phase transition duration and $H_*$ is the Hubble constant at the transition temperature.}
\end{figure}

\begin{figure}[h]
	\centering
	\subfigure{
		\includegraphics[scale=1]{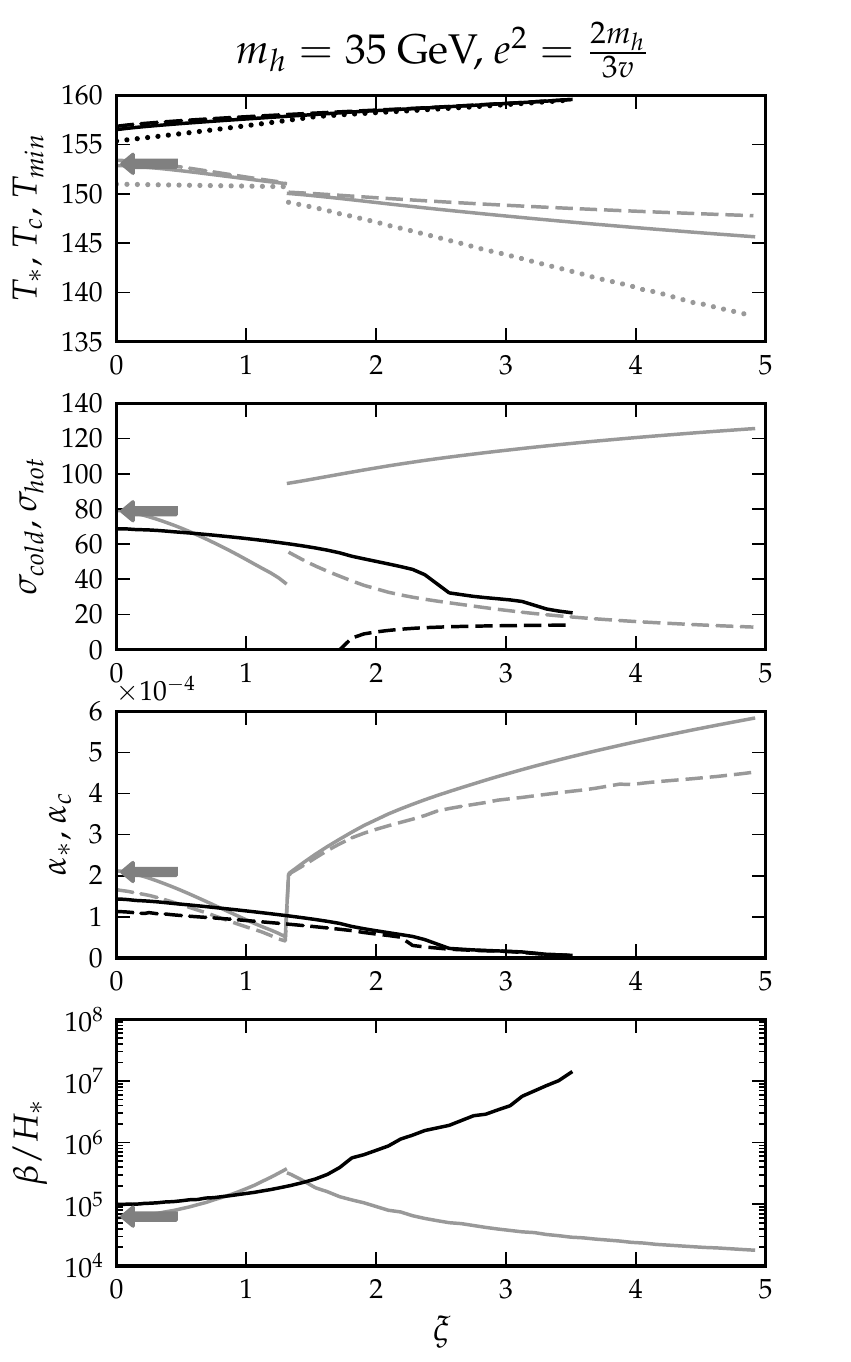}
		\includegraphics[scale=1]{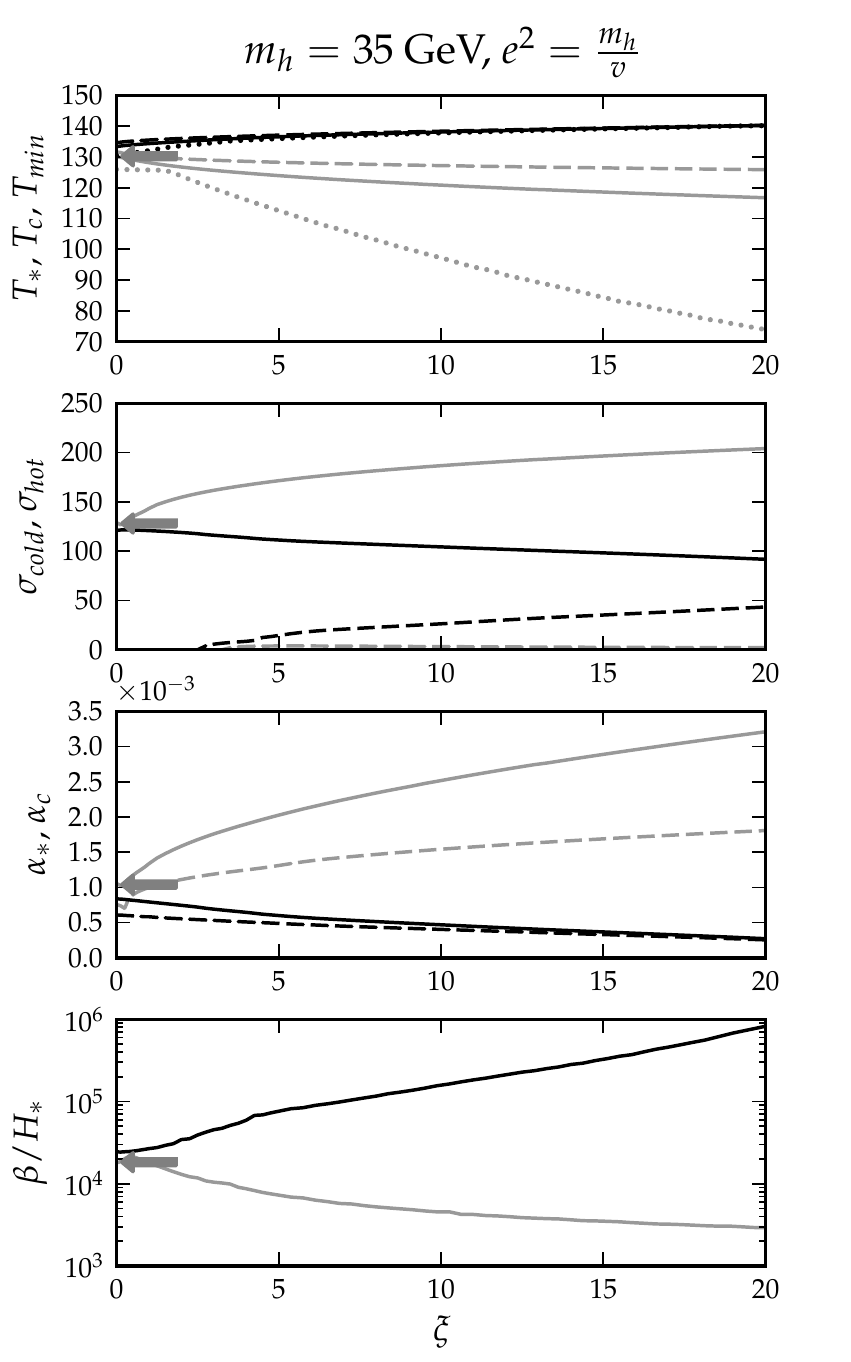}
	}
	\caption{\label{fig:35mh} Calculated gauge dependence of phase transition parameters for a medium-mass Higgs boson. See fig.~\ref{fig:10mh} for a thorough explanation of the different lines. }
\end{figure}

\begin{figure}[h]
	\centering
	\subfigure{
		\includegraphics[scale=1]{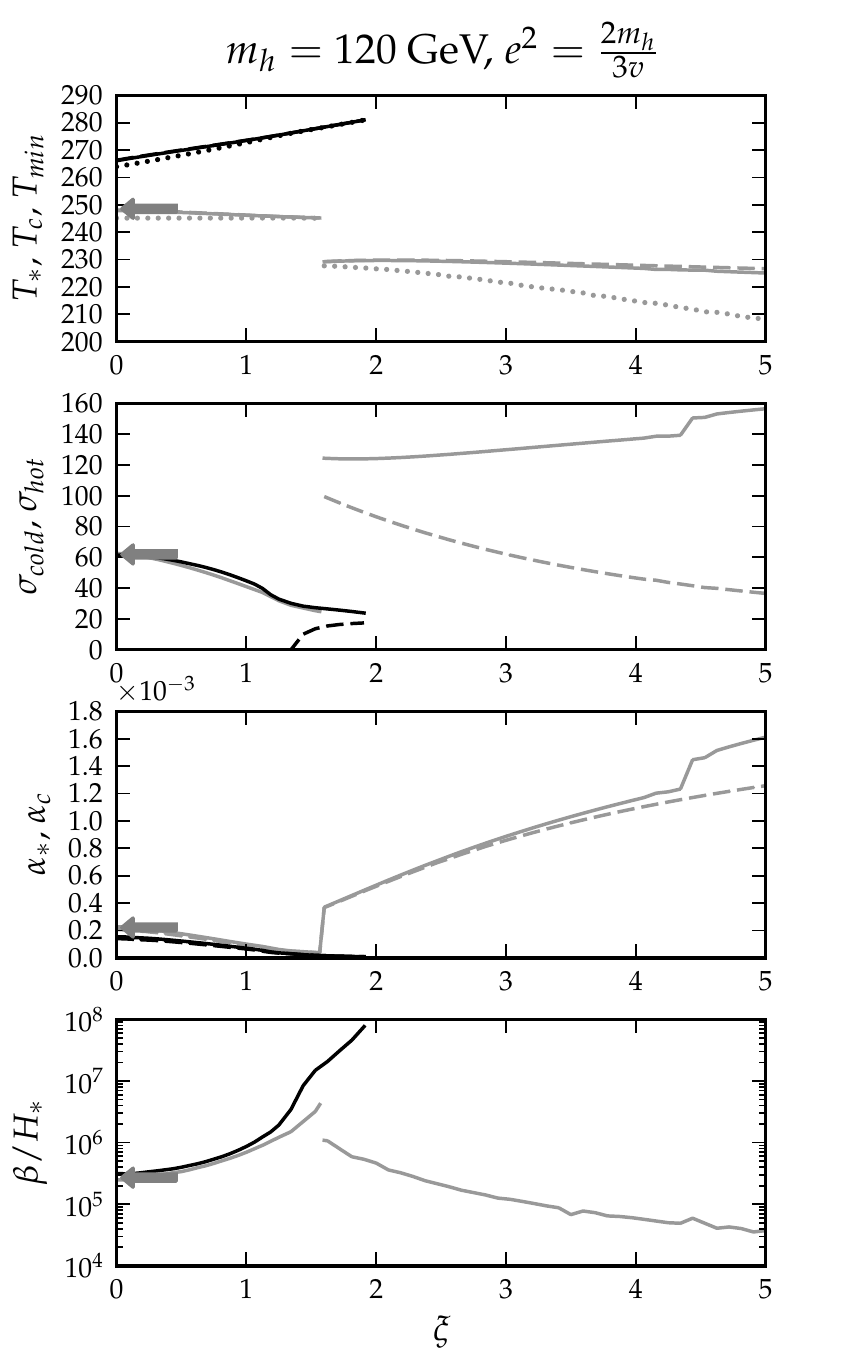}
		\includegraphics[scale=1]{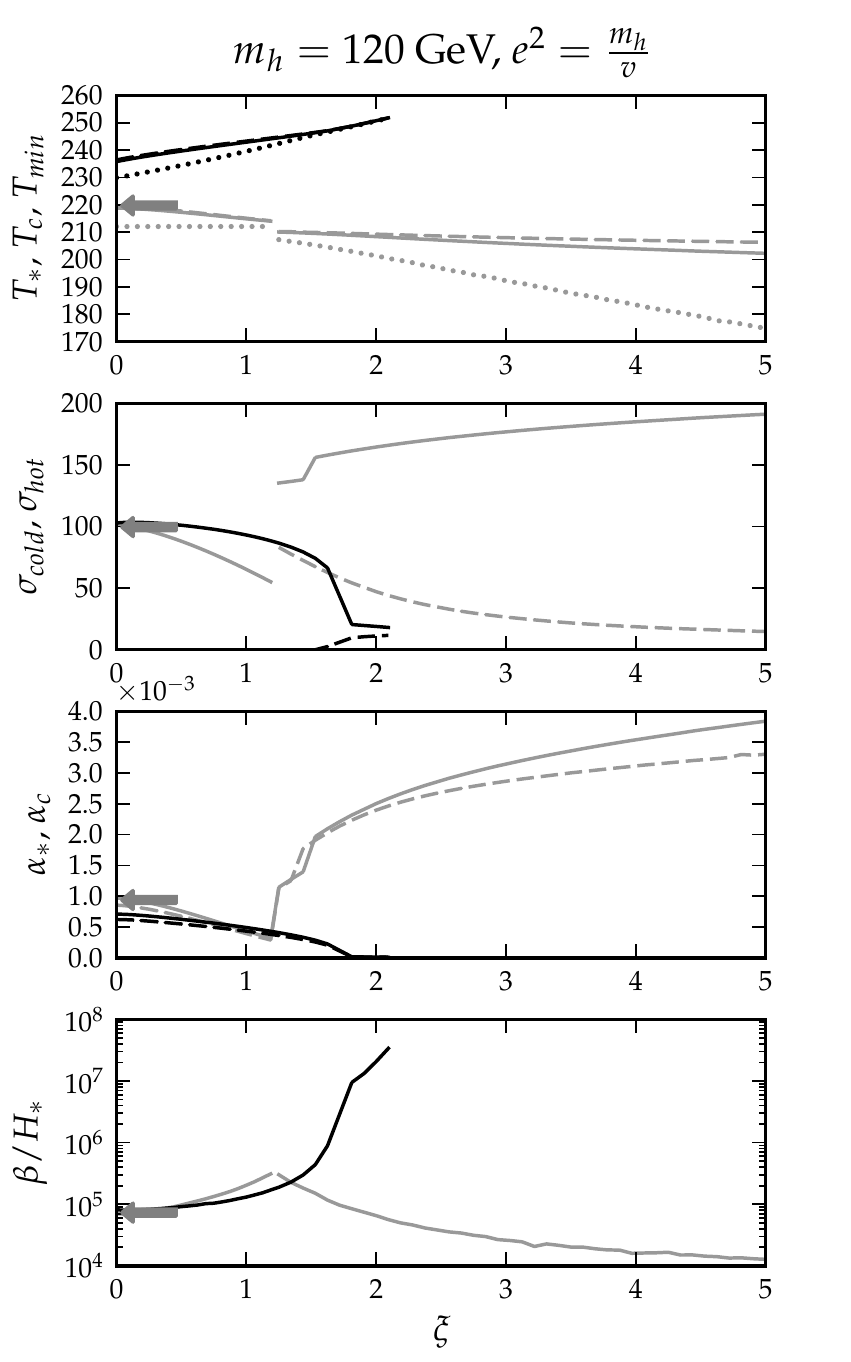}
	}
	\caption{\label{fig:120mh} Calculated gauge dependence of phase transition parameters for a high-mass Higgs boson. See fig.~\ref{fig:10mh} for a thorough explanation of the different lines.}
\end{figure}

Figures \ref{fig:10mh}--\ref{fig:120mh} display our results for the gauge dependence of the different phase transition properties.  All four properties---the transition temperature, the values of $\sigma$ corresponding to the minima of the phases, the relative change in energy density, and the transition duration---heavily depend upon the choice of gauge.  We also show the impact of including the daisy resummation, as discussed above. 

Three broad features emerge from these figures. The results obtained with the Hamiltonian formulation most closely match the results of Landau gauge ($\xi = 0$). However, the match is not exact. Most significantly, the Hamiltonian approach yields a small but measurable increase in the critical and transition temperatures. 

Second, at $\xi=0$, there can exist significant shifts in the GW-wave relevant parameters due to the inclusion of the daisy resummation. Generally, one finds that the values of $\alpha$ are decreased while $\beta/H_\ast$ are increased due to the inclusion of the resummation, implying a reduced amplitude and higher peak frequency in the GW spectrum. This significant departure from the fully gauge-invariant results (albeit within the Landau gauge) suggests that developing a gauge-invariant daisy resummation procedure will be essential for obtaining physically realistic predictions. 

\begin{figure}[t]
	\centering
	\includegraphics[scale=1]{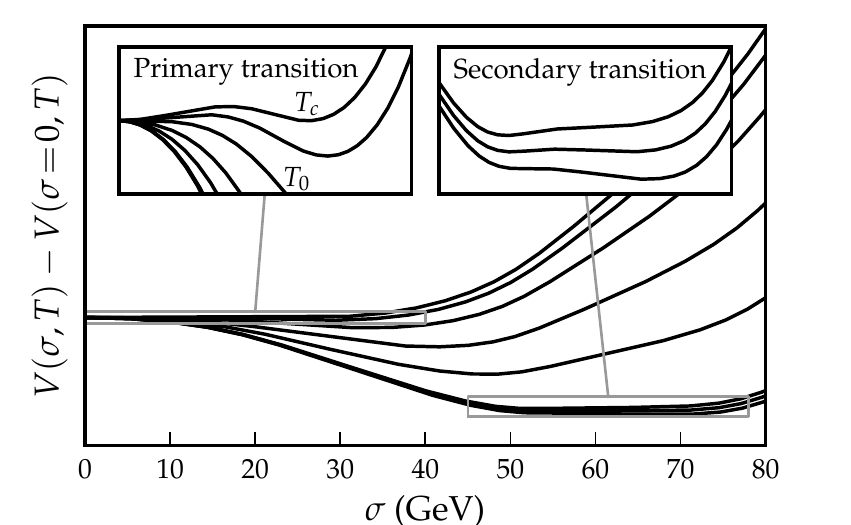}
	\caption{\label{fig:doubleTrans} An example of multiple phase transitions in the same model. Here, $m_h = 35$ GeV, $e^2 = \frac{2m_h}{3v}$, and $\xi = 3$. Since the existence of the secondary phase transition is gauge-dependent, it is clearly non-physical.}
\end{figure}

Third, the dependence on $\xi$ can both exacerbate these differences and lead to new phase structures that are clearly unphysical artifacts of the gauge choice. For example, for $m_h = 120$ GeV (fig.~\ref{fig:120mh}), the phase transition becomes second-order at high $\xi$, so the change in energy density goes to zero and $\beta$ goes to infinity. 
A change in gauge can also lead to secondary minima and secondary transitions (see fig.~\ref{fig:doubleTrans}). In figs.~\ref{fig:10mh}--\ref{fig:120mh}, we always perform calculations for the transition with the largest change in vev, even when this transition happens after initial symmetry breaking. This leads to the discontinuities in figs.~\ref{fig:35mh} and~\ref{fig:120mh}.
Given the unphysical nature of these artifacts, we do not discuss them further but simply point out the danger in this context of attempting to draw physical inferences from a gauge-dependent calculation. 

\begin{figure}[h]
	\centering
	\subfigure{
		\includegraphics[scale=1]{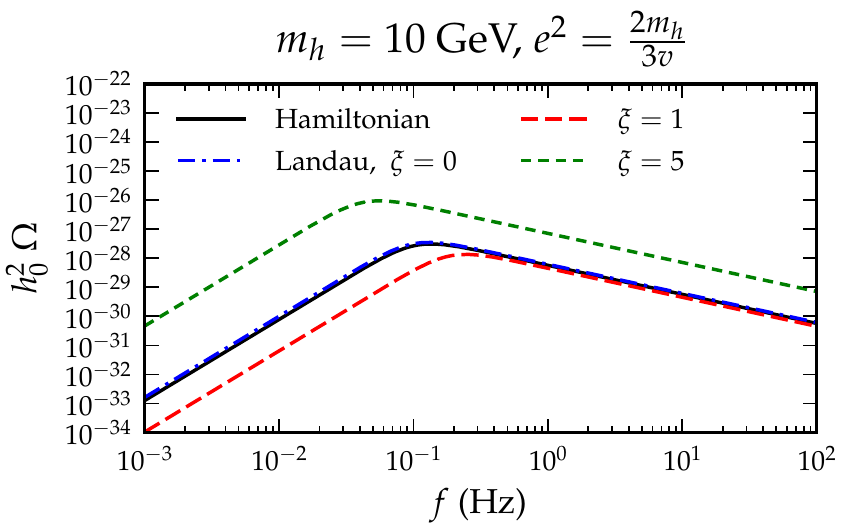}
		\includegraphics[scale=1]{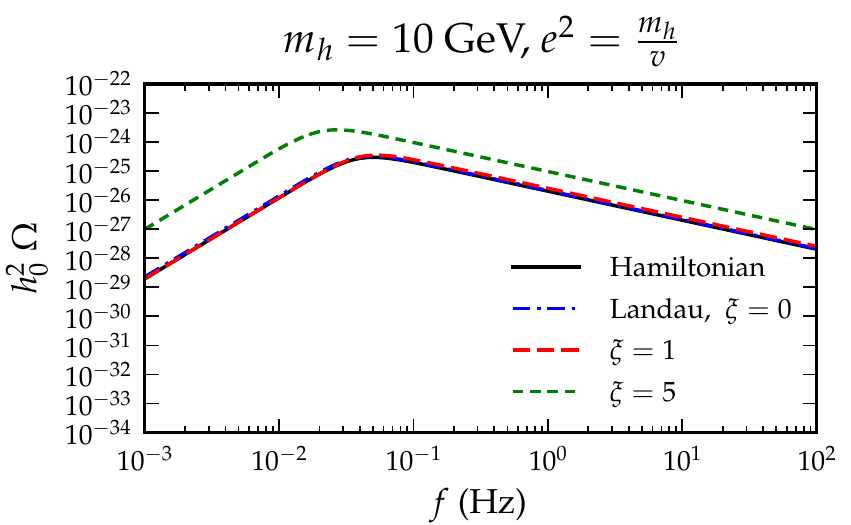}
	}
	\caption{\label{fig:GW10} Expected gravitational wave spectrum for a Higgs mass of 10 GeV, calculated in Landau gauge ($\xi=0$), two high-$\xi$ gauges ($\xi=1,5$), and the gauge-invariant Hamiltonian formalism.}
\end{figure}

\begin{figure}[h]
	\centering
	\subfigure{
		\includegraphics[scale=1]{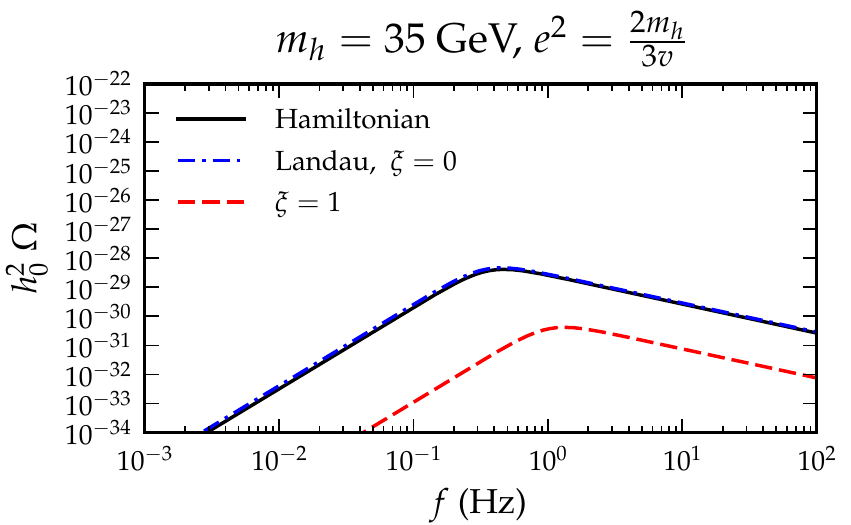}
		\includegraphics[scale=1]{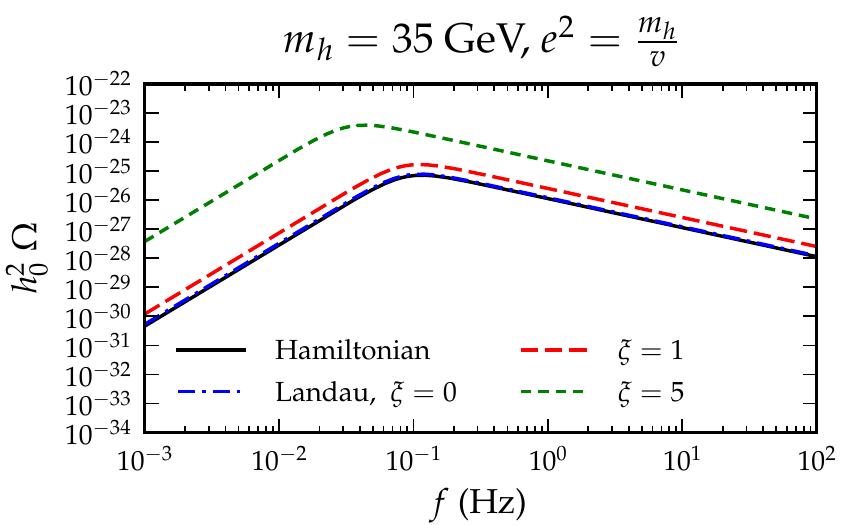}
	}
	\caption{\label{fig:GW35} Expected gravitational wave spectrum for a Higgs mass of 35 GeV, calculated in Landau gauge ($\xi=0$), one or two high-$\xi$ gauges ($\xi=1,5$), and the gauge-invariant Hamiltonian formalism.}
\end{figure}

\begin{figure}[h]
	\centering
	\subfigure{
		\includegraphics[scale=1]{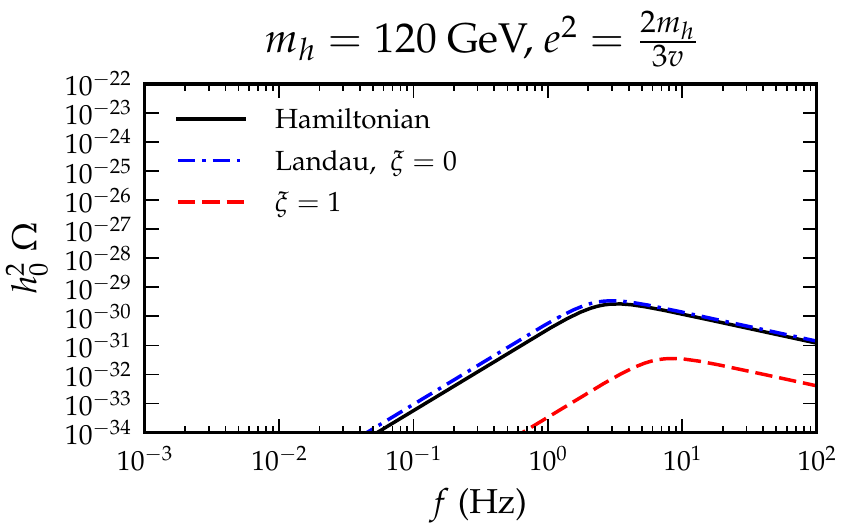}
		\includegraphics[scale=1]{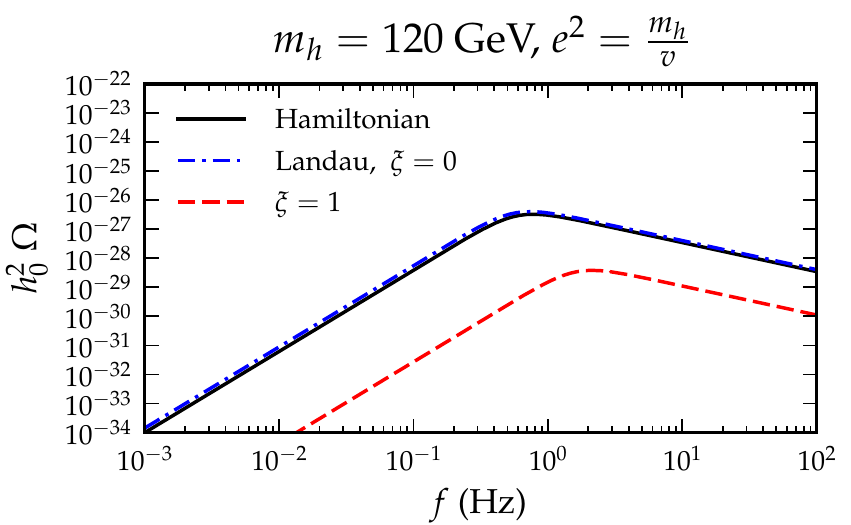}
	}
	\caption{\label{fig:GW120} Expected gravitational wave spectrum for a Higgs mass of 120 GeV, calculated in Landau gauge ($\xi=0$), a high-$\xi$ gauge ($\xi=1$), and the gauge-invariant Hamiltonian formalism.}
\end{figure}

\begin{figure}[h]
	\centering
	\subfigure{
		\includegraphics[scale=1]{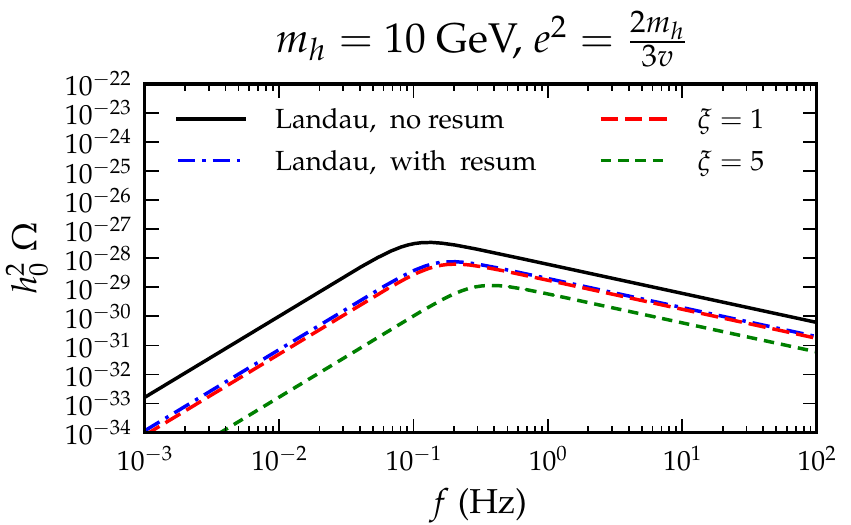}
		\includegraphics[scale=1]{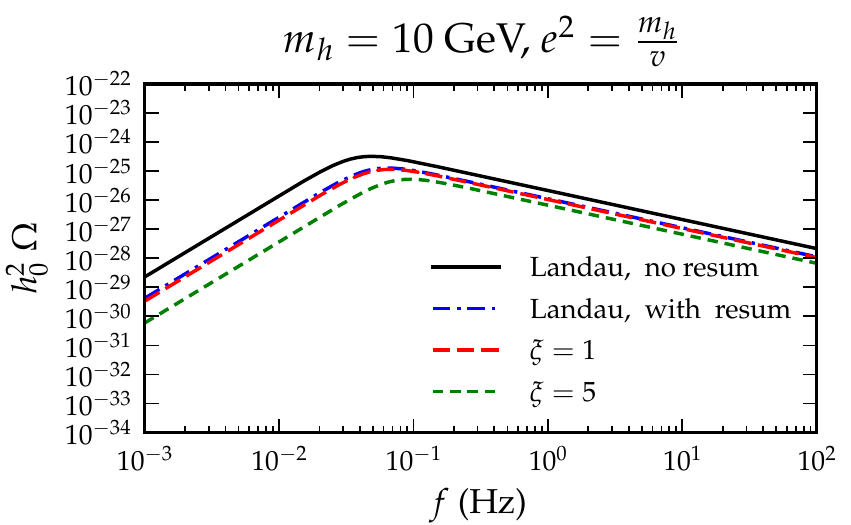}
	}
	\caption{\label{fig:GW10_resum} Comparison of gravitational wave spectra calculated without daisy resummation in Landau gauge, and with resummation (dashed lines) in Landau gauge and two other $R_\xi$ gauges.}
\end{figure}

\begin{figure}[h]
	\centering
	\subfigure{
		\includegraphics[scale=1]{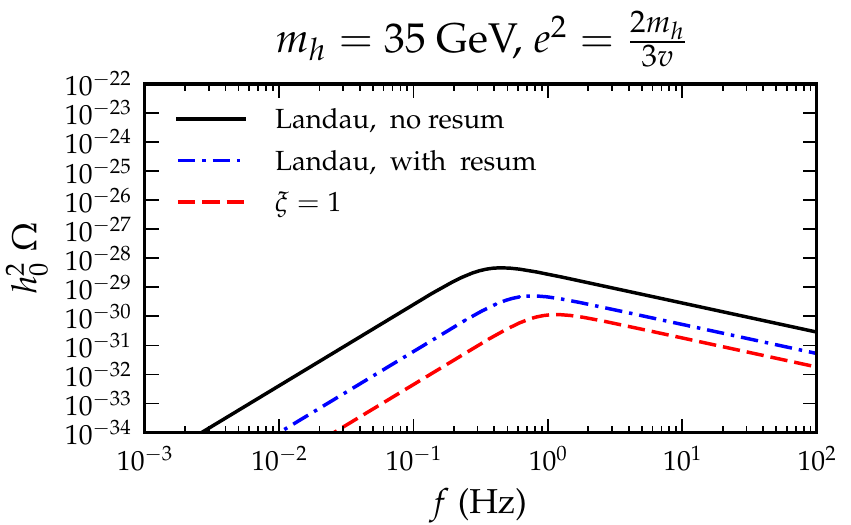}
		\includegraphics[scale=1]{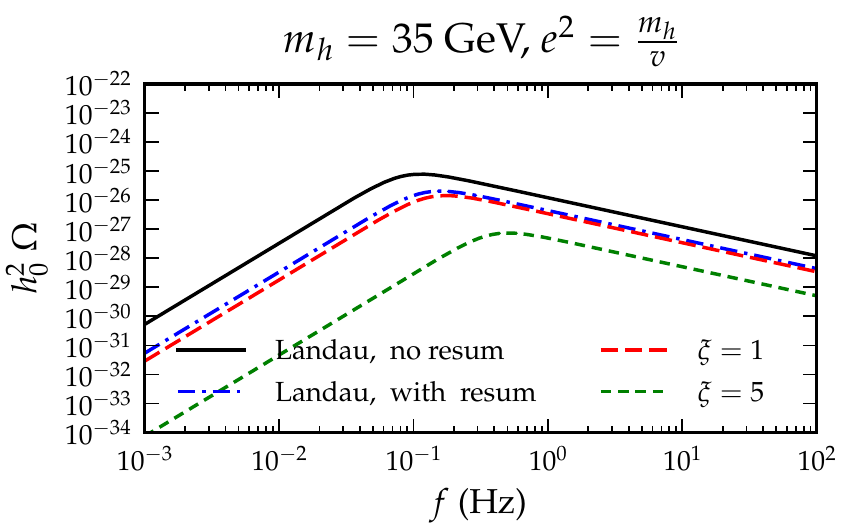}
	}
	\caption {\label{fig:GW35_resum} Comparison of gravitational wave spectra calculated without daisy resummation in Landau gauge, and with resummation (dashed lines) in Landau gauge and one or two other $R_\xi$ gauges.}
\end{figure}

\begin{figure}[h]
	\centering
	\subfigure{
		\includegraphics[scale=1]{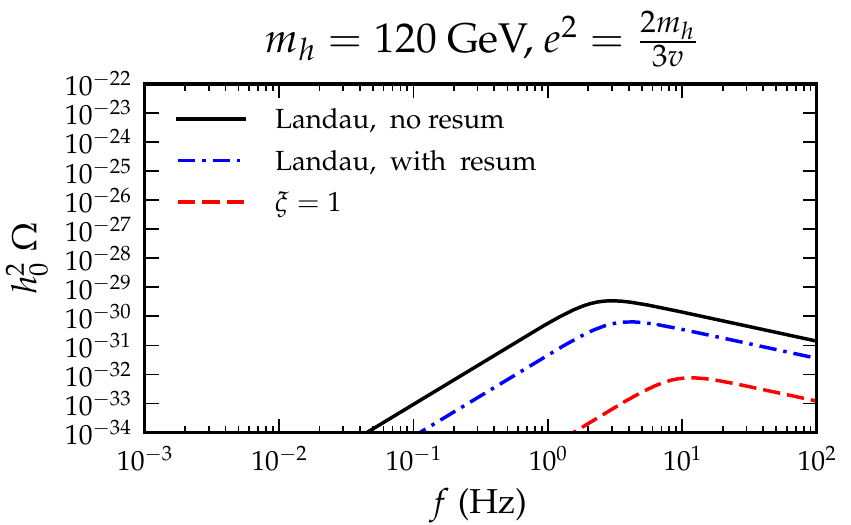}
		\includegraphics[scale=1]{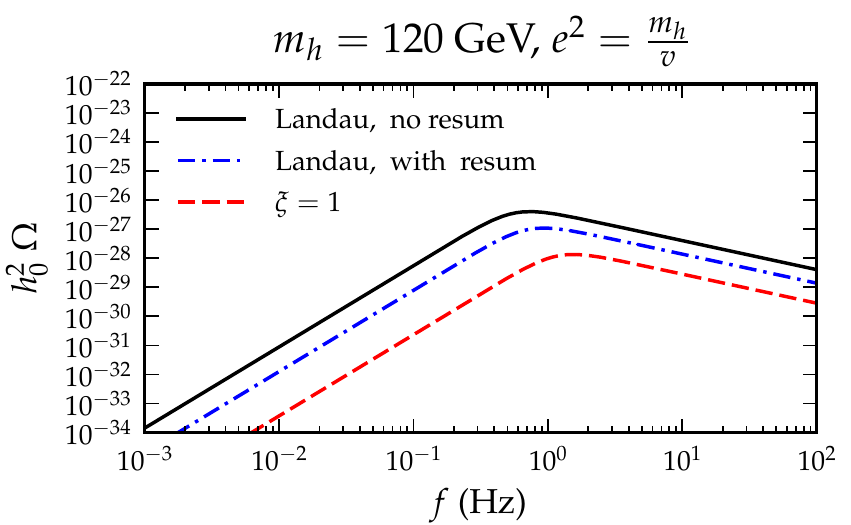}
	}
	\caption {\label{fig:GW120_resum} Comparison of gravitational wave spectra calculated without daisy resummation in Landau gauge, and with resummation (dashed lines) in Landau gauge and one other $R_\xi$ gauge.}
\end{figure}

Finally, we present our calculations for various gravitational wave spectra in figures~\ref{fig:GW10}--\ref{fig:GW120_resum}. 
We make comparisons of the Hamiltonian approach and $R_\xi$ gauges without daisy resummation in figures~\ref{fig:GW10}--\ref{fig:GW120}, and include the effects of resummation in figures~\ref{fig:GW10_resum}--\ref{fig:GW120_resum}.
Again, Landau gauge and the Hamiltonian approach produce very similar results. However, a change in the gauge parameter produces very large changes in both the calculated amplitude and peak frequency of the wave. Without daisy resummation it does not appear possible, or at least feasible, to determine which way this change will manifest without doing the full numerical calculation. With resummation, an increase in the gauge parameter tends to make the phase transition less strongly first-order, thereby decreasing amplitude and increasing the peak frequency of the resulting spectrum.

\section{Discussion}
\label{sec:discussion}
We have thus far presented numerical calculations of strongly first-order phase transitions and spectra of the resultant gravitational waves in the Abelian Higgs model both for various values of the gauge parameter $\xi$ and in two gauge-invariant formalisms. The gauge-invariant Hamiltonian formalism closely matches Landau gauge. We find that small changes in $\xi$ can produce large changes in calculated physical quantities, implying that attention to gauge invariance in GW computations is essential for reaching physically meaningful predictions. Moreover, we find that in the Lagrangian formalism, the result may be significantly affected by inclusion of the daisy resummation, a conclusion similar to what has been observed in the context of sphaleron rate computations\cite{mjrmpatel}. 

From these observations, we conclude that the use of a non-gauge-invariant framework and the neglect of daisy resummations in computations of GW spectra for non-Abelian phase transitions are likely to lead to physically unreliable predictions. At present, it appears that the generalization of gauge-invariant perturbative methods applicable in the Abelian Higgs model to non-Abelian spontaneously broken theories is not straightforward.  The Hamiltonian formalism does not easily carry over to the non-Abelian case and, given how drastically observables change with a change in gauge parameter, one should not trust gauge-dependent calculations for anything other than rough estimates. 

We see several possible directions. First, one can compute thermodynamic quantities of interest  (such as $\alpha$ and $T_\ast$) as well as the bounce action with Monte Carlo methods, thereby circumventing the gauge problem at the outset while including all higher-order effects (including those entering daisy resummed perturbation theory) by construction. The results would undoubtedly be the most reliable theoretically, but this approach is unlikely to be practically feasible for surveying a wide variety of models or exploring wide regions of parameter space for models like the MSSM.  
As an alternative,  following Ref.~\cite{mjrmpatel}, one can use $R_\xi$ gauge and the Nielsen identities to ensure gauge-independence at each order in $\hbar$. The latter approach is relatively straightforward conceptually, but computationally involved, as one must go to at least second order in the loop expansion. It appears particularly challenging in the case of the tunneling rate computation. From a more formal side, it may be possible to construct a gauge-invariant Hamiltonian formalism for spontaneously-broken non-Abelian gauge theories. Although we are not aware of any work in this particular direction, we note that such a formulation has been achieved in the absence of spontaneous symmetry breaking for the specific case of quantum chromodynamics (see, {\em e.g.}, Refs.~\cite{Chen:1996dp,Haller:2003ta} and references therein).

\section*{Acknowledgements}
The authors would like to thank T.~Banks, C.~Caprini, and R.~Holman  for helpful discussions and some very useful references. The authors are also grateful for many useful discussions with H. Patel, particularly in formulating the initial elements of this project.
CW is supported by a National Science Foundation Graduate Fellowship. SP acknowledges support from the National Science Foundation, award PHY-0757911-001, and from an Outstanding Junior Investigator Award from the Department of Energy, DE-FG02-04ER41286. MJRM is supported in part by U.S. Department of Energy Contract DE-FG02-08ER41531 and by the Wisconsin Alumni Research Foundation.

\appendix
\section{Calculating 1-loop zero-T potential}
To calculate the one-loop potential, we must examine integrals like
$\int{\frac{d^3k}{(2\pi)^3}\omega}$.
For $\omega^2 = k^2 + m^2$, this gives the standard one-loop potential associated with a particle of mass $m$. However, in the gauge invariant approach of Boyanovsky et al. the plasma frequency has the form $\omega_p^2 = (k^2+\alpha)(k^2+\beta)/k^2$. They perform the integral using a cutoff regulator, but we would like to use dimensional regularization in order to better compare with the $R_\xi$ gauge.

The potential associated with the plasma mode is given by
\begin{equation}
V_p = \frac{\mu^{3-d}}{2}\int \frac{d^dk}{(2\pi)^d} \left( \frac{k^2}{(k^2+\alpha)(k^2+\beta)} \right)^n
\end{equation}
with $d=3$ and $n=-\frac{1}{2}$, and $\mu$ is a mass dimension that balances the integration measure. 
For $n-\frac{d}{2} > 0$, the integral converges. 
Performing the integral over the d-dimensional sphere yields
\begin{align}
V_p =& \frac{1}{(4\pi)^{d/2}}\frac{1}{\Gamma(d/2)}  \int dk \; k^{d-1}  \left( \frac{k^2}{(k^2+\alpha)(k^2+\beta)} \right)^n \nonumber \\
=& \frac{1}{(4\pi)^{d/2}}\frac{1}{\Gamma(d/2)} \frac{1}{2}  \int dk \; \rho^{d/2-1}  \left( \frac{\rho}{(\rho+\alpha)(\rho+\beta)} \right)^n.
 \end{align}
We can introduce a Feynman parameter to rewrite the fraction as
\begin{equation}
\left( \frac{\rho}{(\rho+\alpha)(\rho+\beta)} \right)^n = \int_0^1 dxdy \; \delta(x+y-1) \frac{ (xy)^{n-1}}{(\rho+\alpha x + \beta y)^{2n}}.
\end{equation}
Using this, and the definition of the beta function
\begin{equation}
\int_0^1 dx\; x^{a-1} (1-x)^{b-1} = B(a,b) = \frac{\Gamma(a)\Gamma(b)}{\Gamma(a+b)},
\end{equation}
one can show that
\begin{equation}
V_p = \frac{1}{(4\pi)^{d/2}} \frac{\Gamma(n+d/2)\Gamma(n-d/2)}{2\Gamma(d/2) \Gamma(n)^2} \int_0^1 dx \; [x(1-x)]^{n-1} [\alpha x  + \beta(1-x)]^{d/2 - n}.
\end{equation}
Then, using the generalized binomial theorem,
\begin{equation}
V_p = \frac{1}{(4\pi)^{d/2}} \frac{\Gamma(n-d/2)}{2\Gamma(d/2) \Gamma(n)^2} \sum^\infty_{l=0}\frac{\Gamma(d/2 - n + 1)\Gamma(d/2-l)\Gamma(n+l)}{\Gamma(d/2-n+1-l)\Gamma(l+1)}  \alpha^{d/2-n-l}\beta^l,
\end{equation}
where we demand that $|\beta|\leq|\alpha|$.
Expanding this out in $\epsilon = \frac{3-d}{2}$, one finds
\begin{multline}
V_p = \frac{1}{64\pi^2} \left[ (\alpha-\beta)^2 \left(-\frac{1}{\epsilon} + \gamma_E - \log(4\pi)  \right) \right. \\
 \left. + (\alpha-\beta)^2 \left(\log\frac{\alpha-\beta}{\mu^2} - \frac{3}{2}  \right) + 4\alpha\beta
\right] + \mathcal{O}(\epsilon),
\end{multline}
where $\gamma_E$ is the Euler-Mascheroni constant. In $\overline{MS}$ regularization, we simply subtract out the term containing $1/\epsilon$, as well as the $\gamma_E$ and $\log(4\pi)$ terms. Note that for $\beta = 0$, this reproduces the standard one-loop potential in equation~\ref{eq:V1}.

\end{document}